\documentclass[11pt]{article}%
\usepackage[slantedGreek]{mathpazo}
\usepackage{amsmath}%
\usepackage{amsfonts}%
\usepackage{amssymb}%
\usepackage{graphicx}
\usepackage{subfigure}
\usepackage{multirow}
\usepackage{booktabs}
\usepackage{setspace}
\usepackage{indentfirst}
\usepackage{hyperref}
\hypersetup{ colorlinks=true,       
    linkcolor=red,          
    citecolor=blue,        
    filecolor=magenta,      
    urlcolor=cyan           
    }

\usepackage{natbib}
\begin{document}

\title{The Market for English Premier League (EPL) Odds\thanks{%
First Draft: April, 2016. We would like to thank the two referees and the associate editor for their valuable comments on the content of our manuscript and their suggestions for improving the document.}}

\author{Guanhao Feng\thanks{%
Address: 5807 S Woodlawn Avenue, Chicago, IL 60637, USA. E-mail address: 
\texttt{guanhao.feng@chicagobooth.edu}.} \\
{\small Booth School of Business}\\
{\small University of Chicago}
\and Nicholas Polson\thanks{%
Address: 5807 S Woodlawn Avenue, Chicago, IL 60637, USA. E-mail address: 
\texttt{nicholas.polson@chicagobooth.edu}.} \\
{\small Booth School of Business}\\
{\small University of Chicago}
\and Jianeng Xu\thanks{%
Address: 5747 S Ellis Avenue, Chicago, IL 60637, USA. E-mail address: 
\texttt{jianeng@uchicago.edu}.} \\
{\small Department of Statistics}\\
{\small University of Chicago}}

\date{January, 2017}
\maketitle

\begin{abstract}
\noindent  This paper employs a Skellam process to represent real-time betting odds for English Premier League (EPL) soccer games. Given a matrix of market odds on all possible score outcomes, we estimate the expected scoring rates for each team. The expected scoring rates then define the implied volatility of an EPL game.  As events in the game evolve, we re-estimate the expected scoring rates and our implied volatility measure to provide a dynamic representation of the market's expectation of the game outcome. Using a dataset of 1520 EPL games from 2012-2016, we show how our model calibrates well to the game outcome. We illustrate our methodology on real-time market odds data for a game between Everton and West Ham in the 2015-2016 season. We show how the implied volatility for the outcome evolves as goals, red cards, and corner kicks occur. Finally, we conclude with directions for future research.
\end{abstract}

\begin{flushleft}
\end{flushleft}

\newpage
\section{Introduction}

\subsection{The betting market for the EPL}
Gambling on soccer is a global industry with revenues between \$700 billion and \$1 trillion a year (see "Football Betting - the Global Gambling Industry worth Billions." BBC Sport). Betting on the result of a soccer match is a rapidly growing market, and online real-time odds exists (Betfair, Bet365, Ladbrokes). Market odds for all possible score outcomes ($0-0, 1-0, 0-1, 2-0, ... $) as well as outright win, lose and draw are available in real time. In this paper, we employ a two-parameter probability model based on a Skellam process and a non-linear objective function to extract the expected scoring rates for each team from the odds matrix. The expected scoring rates then define the implied volatility of the game.

A key feature of our analysis is to use the real-time odds to re-calibrate the expected scoring rates instantaneously as events evolve in the game. This allows us to assess how market expectations change according to exogenous events such as corner kicks, goals, and red cards. A plot of the implied volatility provides a diagnostic tool to show how the market reacts to event information. In particular, we study the evolution of the odds implied final score prediction over the course of the game. Our dynamic Skellam model fits the scoring data well in a calibration study of 1520 EPL games from the 2012 - 2016 seasons. 

The goal of our study is to show how a parsimonious two-parameter model can flexibly model the evolution of the market odds matrix of final scores. We provide a non-linear objective function to fit our Skellam model to instantaneous market odds matrix. We then define the implied volatility of an EPL game and use this as a diagnostics to show how the market's expectation changes over the course of a game. 

One advantage of viewing market odds through the lens of a probability model is the ability to obtain more accurate estimates of winning probabilities. For example, a typical market "vig" (or liquidity premium for bookmakers to make a return) is $5-8\%$ in the win, lose, draw market. Now there is also extra information in the final score odds about the win odds. Our approach helps to extract that information. Another application of the Skellam process is to model final score outcomes as a function of characteristics (see \cite{Karlis:2003ck, Karlis:2009dq}.) 
 
The rest of the paper is outlined as follows. The next subsection provides connections with existing research. Section 2 presents our Skellam process model for representing the difference in goals scored. We then show how to make use of an odds matrix while calibrating the model parameters. We calculate a dynamic implied prediction of any score and hence win, lose and draw outcomes, using real-time online market odds. Section 3 illustrates our methodology using an EPL game between Everton and West Ham during the 2015-2016 season. Finally, Section 4 discusses extensions and concludes with directions for future research.

\subsection{Connections with Existing Work}

There is considerable interest in developing probability models for the evolution of the score of sporting events.
\cite{Stern:1994hj} and \cite{Polson:2015ira} propose a continuous time Brownian motion model for the difference in scores in a sporting event and show how to calculate the implied volatility of a game. 
We build on their approach by using a difference of Poisson processes (a.k.a. Skellam process) for the discrete evolution of the scores of an EPL game, see also 
\cite{Karlis:2003ck, Karlis:2009dq} and \cite{Koopman2014}. 
Early probabilistic models  (\citealt{Lee:1997ct})  predicted the outcome of soccer matches using independent Poisson processes. Later models incorporate a correlation between the two scores and model the number of goals scored by each team using bivariate Poisson models (see \cite{Maher:1982hr} and \cite{Dixon:1997jc}). Our approach follows \cite{Stern:1994hj} by modeling the score difference (a.k.a. margin of victory),  instead of modeling the number of goals and the correlation between scores directly. 

There is also an extensive literature on soccer gambling and market efficiency. For example, \cite{Vecer2009} estimates the scoring intensity in a soccer game from betting markets. \cite{Dixon:2004gj} presents a detailed comparison of odds set by different bookmakers. \cite{Fitt:2009iv} uses market efficiency to analyze the mispricing of cross-sectional odds
and \cite{Fitt:2005bj} models online soccer spread bets.

Another line of research, asks whether betting markets are efficient and, if not, how to exploit potential inefficiencies in the betting market. For example, \cite{Levitt2004} discusses the structural difference of the gambling market and financial markets. The study examines whether bookmakers are more skilled at game prediction than bettors and in turn exploit bettor biases by setting prices that deviate from the market clearing price.  \cite{Avery:1999jg} examine the hypothesis that sentimental bettors act like noise traders and can affect the path of prices in soccer betting markets. 

\section{Skellam Process for EPL scores}

To model the outcome of a soccer game between team A and team B, we let the difference in scores, $N(t)=N_A(t)-N_B(t)$ where
 $N_A(t)$ and $N_B(t)$ are the team scores at time point $t$. Negative values of $N(t)$ indicate that team A is behind. We  begin at  $N(0) = 0$ and ends at time one with $N(1)$ representing the final score difference. The probability $\mathbb{P}(N(1)>0)$ represents the ex-ante odds of team A winning. 
Half-time score betting, which is common in Europe, is available for the distribution of $N(\frac{1}{2})$. 

We develop a probabilistic model for the distribution of $N(1)$ given $N(t)=\ell$ where $\ell$ is the current lead. This model, together with the current market odds can be used to infer the expected scoring rates of the two teams and then to define the implied volatility of the outcome of the match.  We let $ \lambda^A$ and $ \lambda^B $ denote the expected scoring rates for the whole game. We allow for the possibility that the scoring abilities (and their market expectations) are time-varying, in which case we denote the expected scoring rates after time $t$ by $ \lambda^A_t $ and $\lambda^B_t$ respectively, instead of $ \lambda^A(1-t) $ and $\lambda^B(1-t)$.

\subsection{Implied  Score Prediction from EPL Odds}
The Skellam distribution is defined as the difference between two independent Poisson variables, see \cite{Skellam:1946kb}, \cite{Sellers:2012uy}, \cite{Alzaid:2010ua}, and \cite{BarndorffNielsen:2012tx}. \cite{Karlis:2009dq} shows how Skellam distribution can be extended to a difference of distributions which have a specific trivariate latent variable structure.
Following \cite{Karlis:2003ck}, we decompose the scores of each team as
\begin{equation}
\left\{
\begin{aligned}
N_A(t) &=& W_A(t)+W(t) \\
N_B(t) &=& W_B(t)+W(t)
\end{aligned}
\right.
\end{equation}
where $W_A(t)$, $W_B(t)$ and $W(t)$ are independent processes with 
$W_A(t) \sim Poisson (\lambda^A  t)$, $W_B(t) \sim Poisson (\lambda^B  t) . $
Here $W(t)$ is a non-negative integer-valued process to induce a correlation between the numbers of goals scored. 
By modeling the score difference, $N(t)$, we avoid having to specify the distribution of $W(t)$ as the difference in goals scored is independent of $W(t)$. Specifically, we have
a Skellam distribution 
\begin{equation}
N(t) = N_A(t) - N_B(t) = W_A(t) - W_B(t) \sim Skellam(\lambda^A  t,\lambda^B  t).
\label{skellam}
\end{equation}
where $ \lambda^A t $ is the cumulative expected scoring rate on the interval $ [0,t]$.
At time $t$, we have the conditional distributions 
\begin{equation}
\left\{
\begin{aligned}
W_A(1) - W_A(t) &\sim& Poisson (\lambda^A(1-t)) \\
W_B(1) - W_B(t) &\sim& Poisson (\lambda^B(1-t)) \\
\end{aligned}
\right.
\end{equation}
Now letting $N^*(1-t)$, the score difference of the sub-game which starts at time $t$ and ends at time 1 and the duration is $(1-t)$. By construction, $N(1) = N(t) + N^*(1-t)$. Since $N^*(1-t)$ and $N(t)$ are differences of two Poisson process on two disjoint time periods, by the property of Poisson process, $N^*(1-t)$ and $N(t)$ are independent. 
Hence, we can re-express equation (\ref{skellam}) in terms of $N^*(1-t)$, and deduce
\begin{equation}
N^*(1-t) = W^*_A(1-t) - W^*_B(1-t) \sim Skellam(\lambda^A_t,\lambda^B_t)
\end{equation}
where $W^*_A(1-t) = W_A(1) - W_A(t)$, $\lambda^A = \lambda^A_0$ and $\lambda^A_t=\lambda^A(1-t)$. A natural interpretation of the expected scoring rates, $\lambda^A_t$ and $\lambda^B_t$, is that they reflect the "net" scoring ability of each team from time $t$ to the end of the game.  The term $W(t)$
model a common strength due to external factors, such as weather. The "net" scoring abilities of the two teams are assumed to be independent of each other as well as the common strength factor. 
We can calculate the probability of any particular score difference, given by $\mathbb{P}(N(1)=x|\lambda^A,\lambda^B)$, at the end of the game where the $ \lambda$'s are estimated from the matrix of market odds. Team strength and "net" scoring ability can be influenced by various underlying factors, such as the offensive and defensive abilities of the two teams. The goal of our analysis is to only represent these parameters at every instant as a function of the market odds matrix for all scores.

To derive the implied winning probability, we use the law of total probability. The probability mass function of a Skellam random variable is the convolution of two Poisson distributions:
\begin{eqnarray}
\mathbb{P}(N(1)=x|\lambda^A,\lambda^B)
&=&\sum_{k=0}^\infty \mathbb{P}(W_B(1)=k-x|W_A(1)=k, \lambda^B)  \mathbb{P}(W_A(1)=k|\lambda^A) \nonumber\\
&=&\sum_{k=max\{0,x\}}^\infty \left\{e^{-\lambda^B}\frac{(\lambda^B)^{k-x}}{(k-x)!}\right\}\left\{e^{-\lambda^A}\frac{(\lambda^A)^k}{k!}\right\}\nonumber\\
&=&e^{-(\lambda^A+\lambda^B)} \sum_{k=max\{0,x\}}^\infty\frac{(\lambda^B)^{k-x}(\lambda^A)^k}{(k-x)!k!}\nonumber \\
&=&e^{-(\lambda^A+\lambda^B)} \left(\frac{\lambda^A}{\lambda^B}\right)^{x/2}I_{|x|}(2\sqrt{\lambda^A\lambda^B})
\end{eqnarray}
where $I_r(x)$ is the modified Bessel function of the first kind (for full details, see \cite{Alzaid:2010ua}), thus has the series representation
\[ I_r(x)=\left(\frac{x}{2}\right)^r \sum_{k=0}^{\infty} \frac{(x^2/4)^k}{k!\Gamma(r+k+1)}. \]
The probability of home team A winning is given by 
\begin{equation}
 \mathbb{P}(N(1)>0|\lambda^A,\lambda^B)=\sum_{x=1}^\infty \mathbb{P}(N(1)=x|\lambda^A,\lambda^B). 
\end{equation}
In practice, we truncate the number of possible goals since the probability of an extreme score difference is negligible. Unlike the Brownian motion model for the evolution of the outcome in a sports game (\cite{Stern:1994hj}, \cite{Polson:2015ira}), the probability of a draw in our setting is not zero. Instead, $\mathbb{P}(N(1)=0|\lambda^A,\lambda^B)>0$ depends on the sum and product of two parameters $\lambda^A$ and $\lambda^B$ and thus the odds of a draw are non-zero. 

\begin{figure}[ht!]
\centering 
\includegraphics[scale=0.5]{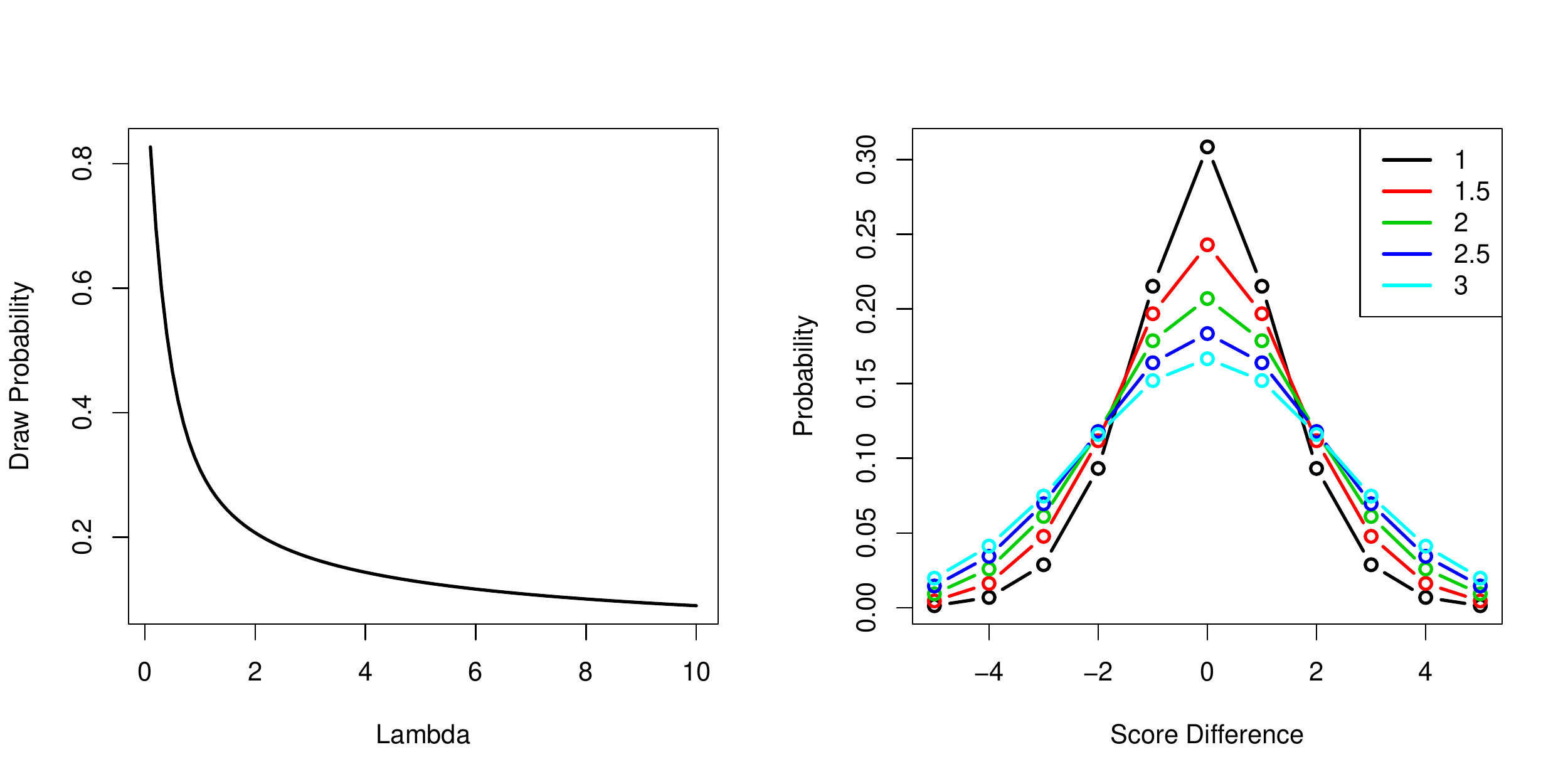} 
\caption{Left: Probability of a draw for two evenly matched teams. Right: Probability of score differences for two evenly matched teams. Lambda values are denoted by different colors.}
\label{draw}
\end{figure}

For two evenly matched teams with$\lambda^A=\lambda^B=\lambda$, we have
\begin{equation}
\mathbb{P}(N(1)=0|\lambda^A=\lambda^B=\lambda) 
= e^{-2\lambda}I_0(2\lambda) 
= \sum_{k=0}^{\infty} \frac{1}{(k!)^2}\left(\frac{\lambda^k}{e^\lambda}\right)^2.
\end{equation}
Figure \ref{draw} shows that this probability is a monotone decreasing function of $\lambda$  and so two evenly matched teams with large $\lambda$'s are less likely to achieve a draw.

Another quantity of interest is the conditional probability of winning as the game progresses. If the current lead at time $t$ is $\ell$, and $N(t)=\ell=N_A(t)-N_B(t)$,
the Poisson property implied that the final score difference $(N(1)|N(t)=\ell)$can be calculated by using the fact that  $N(1)=N(t)+N^*(1-t)$ and $N(t)$ and $N^*(1-t)$ are independent. Specifically, conditioning on $N(t)=\ell$, we have the identity 
\[ N(1)=N(t)+N^*(1-t)=\ell+Skellam(\lambda^A_t,\lambda^B_t).\] 

We are now in a position to find the conditional distribution ($N(1)=x|N(t)=\ell$) for every time point $t$ of the game given the current score. Simply put, we have the time homogeneous condition
\begin{eqnarray}
\mathbb{P}(N(1)=x|\lambda^A_t,\lambda^B_t,N(t)=\ell)&=&\mathbb{P}(N(1)-N(t)=x-\ell |\lambda^A_t,\lambda^B_t,N(t)=\ell)\nonumber\\
&=&\mathbb{P}(N^* (1-t)=x-\ell |\lambda^A_t,\lambda^B_t)
\end{eqnarray}
where $\lambda^A_t$, $\lambda^B_t$, $\ell$ are given by market expectations at time $t$. 

Two conditional probabilities of interest are he chances that the home team A wins, 
\begin{eqnarray}
\mathbb{P}(N(1)>0|\lambda^A_t,\lambda^B_t,N(t)=\ell)&=&\mathbb{P}(\ell+ N^*(1-t)>0|\lambda^A_t,\lambda^B_t)\nonumber\\
&=&\mathbb{P}(Skellam(\lambda^A_t,\lambda^B_t)>-\ell |\lambda^A_t,\lambda^B_t)\nonumber\\
&=&\sum_{x>-\ell}e^{-(\lambda^A_t+\lambda^B_t)}\left(\frac{\lambda^A_t}{\lambda^B_t}\right)^{x/2}I_{|x|}(2\sqrt{\lambda^A_t\lambda^B_t}).
\end{eqnarray}
and the conditional probability of a draw at time $t$ is 
\begin{eqnarray}
\mathbb{P}(N(1)=0|\lambda^A_t,\lambda^B_t,N(t)=\ell)&=&\mathbb{P}(\ell+N^*(1-t)=0|\lambda^A_t,\lambda^B_t)\nonumber\\
&=&\mathbb{P}(Skellam(\lambda^A_t,\lambda^B_t)=-\ell |\lambda^A_t,\lambda^B_t)\nonumber\\
&=&e^{-(\lambda^A_t+\lambda^B_t)}\left(\frac{\lambda^A_t}{\lambda^B_t}\right)^{-\ell/2}I_{|\ell |}(2\sqrt{\lambda^A_t\lambda^B_t}).
\end{eqnarray}

\noindent The conditional probability at time $t$ of home team A losing is 
$ 1-\mathbb{P}(N(1)>0|\lambda^A_t,\lambda^B_t,N(t)=\ell) $.
We now turn to the calibration of our model from given market odds.

\subsection{Market Calibration}
Our information set at time $t$, denoted by $\mathcal{I}_t$, includes the current lead $N(t) = \ell$ and the market odds for $\left\{Win, Lose, Draw, Score\right\}_t$, where 
$Score_t = \{ ( i - j ) : i, j = 0, 1, 2, ....\}$. These market odds can be used to calibrate a Skellam distribution which has only two parameters $\lambda^A_t$ and $\lambda^B_t$. The best fitting Skellam model with parameters $\{\hat\lambda^A_t,\hat\lambda^B_t\}$ will then provide a better estimate of the market's information concerning the outcome of the game than any individual market (such as win odds) as they are subject to a "vig" and liquidity. Suppose that the fractional odds for all possible final score outcomes are given by a bookmaker. In this case, the bookmaker pays out three times the amount staked by the bettor if the outcome is indeed 2-1. Fractional odds are used in the UK, while money-line odds are favored by American bookmakers with $2:1$ ("two-to-one") implying that the bettor stands to make a \$200 profit on a \$100 stake. The market implied probability makes the expected winning amount of a bet equal to 0. In this case, the implied probability $p=1/(1+3)=1/4$ and the expected winning amount is $\mu=-1*(1-1/4)+3*(1/4)=0$. We denote this odds as $odds(2,1)=3$. To convert all the available odds to  implied probabilities, we use the identity 
\[ \mathbb{P}(N_A(1) = i, N_B(1) = j)=\frac{1}{1+odds(i,j)}. \]
The market odds matrix, $O$, with elements $o_{ij}=odds(i-1,j-1)$, $i,j=1,2,3...$ provides all possible combinations of final scores. Odds on extreme outcomes are not offered by the bookmakers. Since the probabilities are tiny, we set them equal to 0. The sum of the possible probabilities is still larger than 1  (see \cite{Dixon:1997jc} and \cite{Polson:2015ira}). This "excess" probability corresponds to a quantity known as the "market vig." For example, if the sum of all the implied probabilities is 1.1, then the expected profit of the bookmaker is 10\%. To account for this phenomenon, we scale the probabilities to sum to 1 before estimation. 

To estimate the expected scoring rates,  $\lambda^A_t$ and $\lambda^B_t$, for the sub-game $N^*(1-t)$, the odds from a bookmaker should be adjusted by $N_A(t)$ and $N_B(t)$. For example, if $N_A(0.5)=1$, $N_B(0.5)=0$ and $odds(2,1)=3$ at half time, these observations actually says that the odds for the second half score being 1-1 is 3 (the outcomes for the whole game and the first half are 2-1 and 1-0 respectively, thus the outcome for the second half is 1-1). The adjusted ${odds}^*$ for $N^*(1-t)$ is calculated using the original odds as well as the current scores and given by 
\begin{equation}
{odds}^*(x,y)=odds(x+N_A(t),y+N_B(t)).
\end{equation}

At time $t$ $(0\leq t\leq 1)$, we calculate the implied conditional probabilities of score differences using odds information
\begin{equation}
\mathbb{P}(N(1)=k|N(t)=\ell)=\mathbb{P}(N^*(1-t)=k-\ell)=\frac{1}{c}\sum_{i-j=k-\ell}\frac{1}{1+{odds}^*(i,j)}\end{equation}
where $c=\sum_{i,j} \frac{1}{1+{odds}^*(i,j)}$ is a scale factor, $\ell=N_A(t)-N_B(t)$, $i,j\geq 0$ and $k=0,\pm 1,\pm 2\ldots$.

Moments of the Poisson distribution make it straightforward to derive the moments of a Skellam random variable with parameters $\lambda^A$ and $\lambda^B$. The unconditional mean and variance are given by $$E[N(1)]=E[W_A(1)]-E[W_B(1)]=\lambda^A-\lambda^B,$$ 
$$V[N(1)]=V[W_A(1)]+V[W_B(1)]=\lambda^A+\lambda^B.$$ Therefore, the conditional moments are given by 
\begin{equation}
\left\{
\begin{aligned}
E[N(1)|N(t)=\ell]&=\ell+(\lambda^A_t-\lambda^B_t),\\
V[N(1)|N(t)=\ell]&=\lambda^A_t+\lambda^B_t.
\end{aligned}
\right.
\end{equation}
We also need to ensure that  $\hat E[N(1)|N(t)=\ell]-\ell\leq \hat V[N(1)|N(t)=\ell]$. A method of moments estimate of $\lambda$'s is given by the  solution to
\begin{equation}
\left\{
\begin{aligned}
\hat E[N(1)|N(t)=\ell]&=\ell+(\lambda^A_t-\lambda^B_t),\\
\hat V[N(1)|N(t)=\ell]&=\lambda^A_t+\lambda^B_t,
\end{aligned}
\right.
\end{equation}
where $\hat E$ and $\hat V$ are the expectation and variance calculated using market implied conditional probabilities, could be negative. To address this issue, we define the residuals
\begin{equation}
\left\{
\begin{aligned}
D_E&=\hat E[N(1)|N(t)=\ell]-[\ell+(\lambda^A_t-\lambda^B_t)],\\
D_V&=\hat V[N(1)|N(t)=\ell]-(\lambda^A_t+\lambda^B_t).
\end{aligned}
\right.
\end{equation}
We then calibrate parameters by adding the constraints $\lambda^A_t\geq 0$ and $\lambda^B_t\geq 0$ and solving the following equivalent constrained optimization problem. 
\begin{eqnarray}
\left(\hat\lambda^A_t,\hat\lambda^B_t\right) &=& \underset{\lambda^A_t,\lambda^B_t}{\arg\min} \left\{D_E^2+D_V^2\right\}\\
&\text{subject to} & \lambda^A_t\geq 0, \lambda^B_t\geq 0 \nonumber
\end{eqnarray}

\begin{figure}[ht!]
\centering 
\includegraphics[scale=0.5]{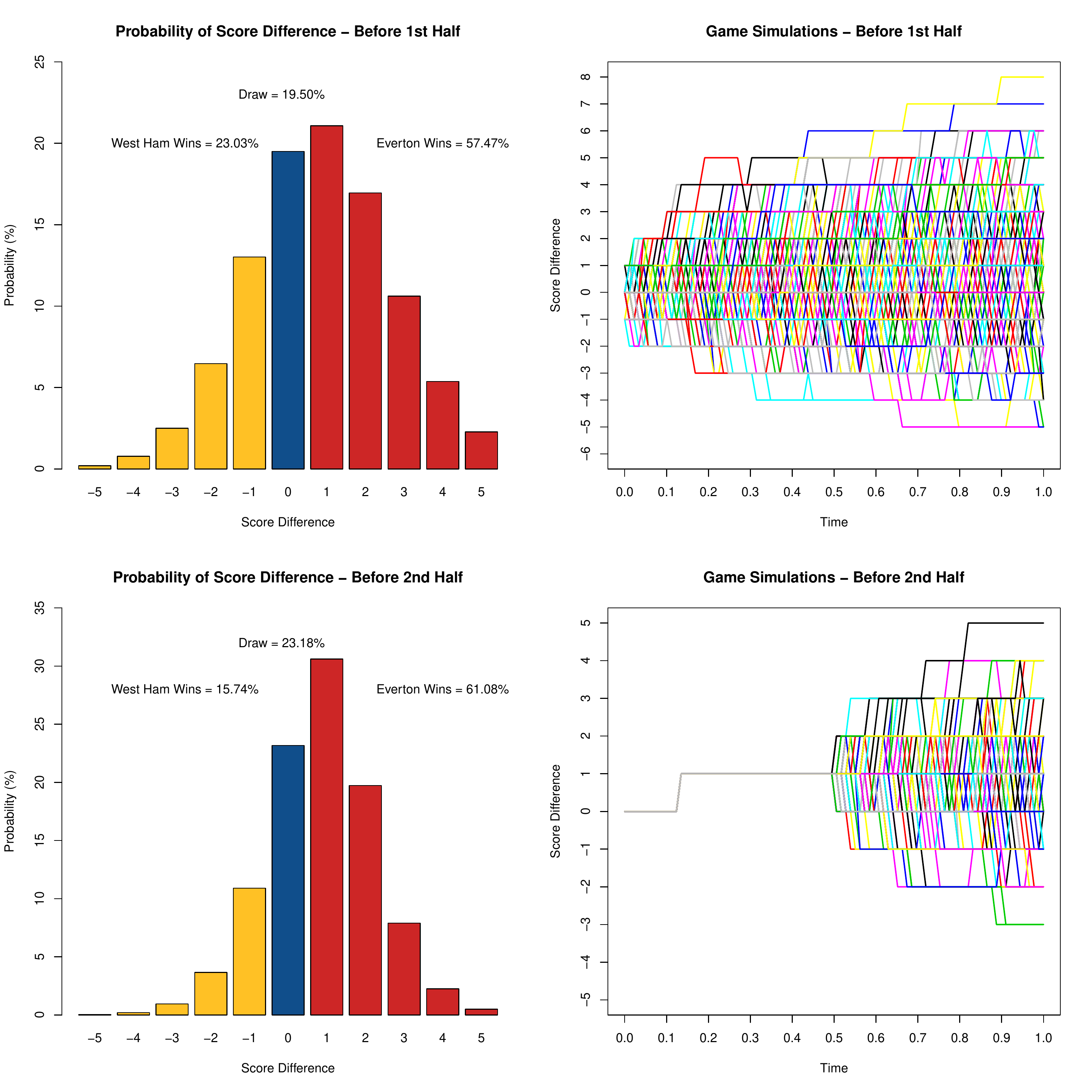} 
\caption{The Skellam process model for winning margin and game simulations. The top left panel shows the outcome distribution using odds data before the match starts. Each bar represents the probability of a distinct final score difference, with its color corresponding to the result of win/lose/draw. Score differences larger than 5 or smaller than -5 are not shown. The top right panel shows a set of simulated Skellam process paths for the game outcome. The bottom row has the two figures updated using odds data available at half-time.} 
\label{prob}
\end{figure}

Figure \ref{prob} illustrates a simulation evolution of an EPL game between Everton and West Ham (March 5th, 2016) with their estimated parameters. It provides a discretized version of Figure 1 in \cite{Polson:2015ira}. The outcome probability of first half and updated second half are given in the left two panels. The top right panel illustrates a simulation-based approach to visualizing how the model works in the dynamic evolution of score difference. In the bottom left panel, from half-time onwards, we also simulate a set of possible Monte Carlo paths to the end of the game. This illustrates the discrete nature of our Skellam process and how the scores evolve.

\subsection{Model Diagnostics}
\begin{figure}[ht!]
\centering 
\includegraphics[scale=0.5]{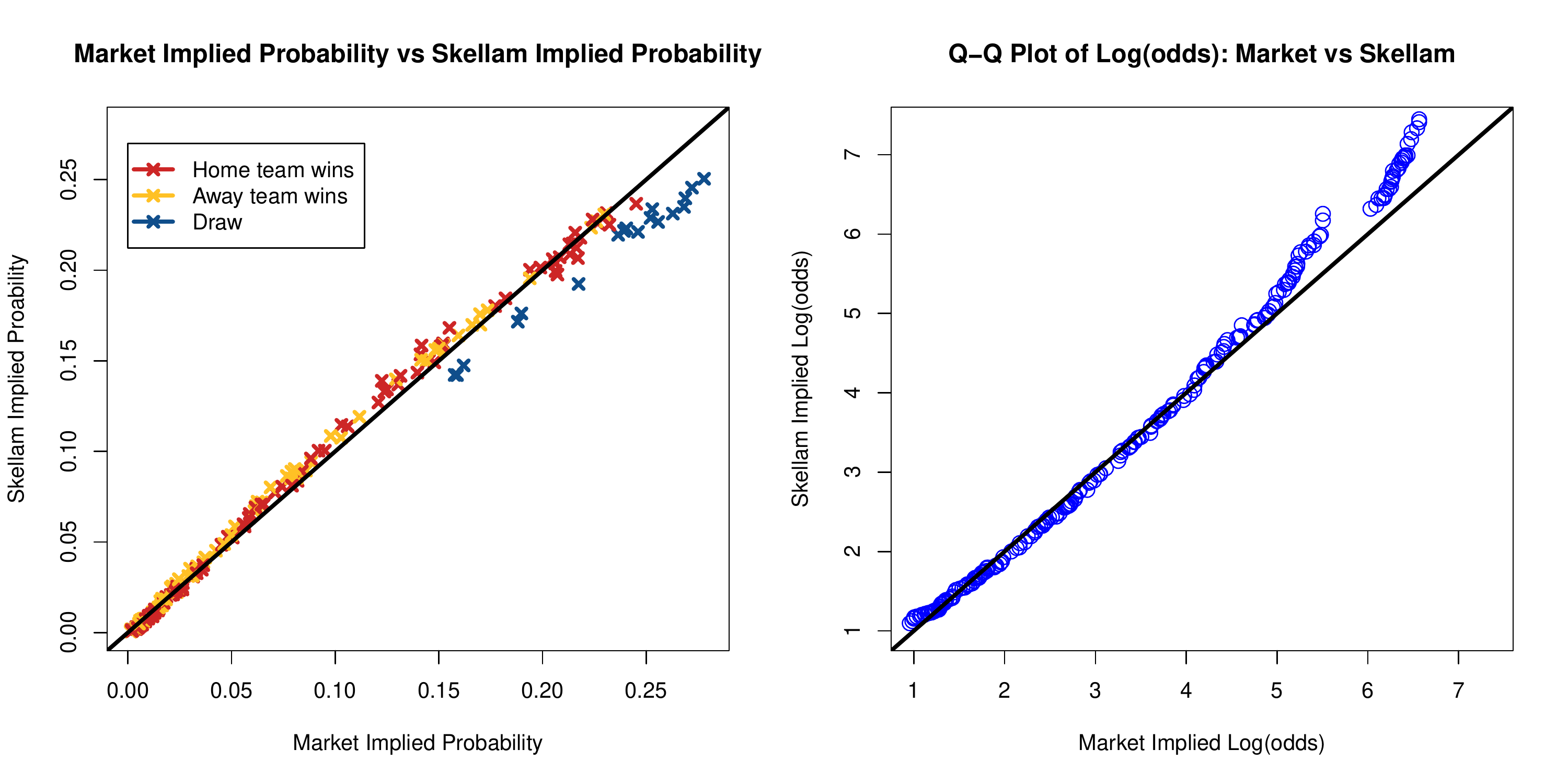} 
\caption{Left: Market implied probabilities for the score differences versus Skellam implied probabilities. Every data point represents a particular score difference; Right: Market log(odds) quantiles versus Skellam implied log(odds) quantiles. Market odds (from  {\tt ladbrokes.com}) of 18 games in EPL 2016-2017 are used (in average 13 score differences per game). The total number of outcomes is 238.}
\label{18games}
\end{figure}

To assess the performance our score-difference Skellam model calibration for the market odds, we have collected data from {\tt ladbrokes.com} on the correct score odds of 18 EPL games (from October 15th to October 22nd, 2016) and plot the calibration result in Figure \ref{18games}. The Q-Q plot of $\log(odds)$ is also shown. In average, there are 13 different outcomes per game, i.e., $N(1) = -6, -5, ... 0, ..., 5, 6$. In total 238 different outcomes are used. We compare our Skellam implied probabilities with the market implied probabilities for every outcome of the 18 games. If the model calibration is sufficient, all the data points should lies on the diagonal line. 
Figure \ref{18games} left panel demonstrates that our Skellam model is calibrated by the market odds sufficiently well, except for the underestimated draw probabilities. \cite{Karlis:2009dq} describe this underestimation phenomenon in a Poisson-based model for the number of goals scored. Following their approach, we apply a zero-inflated version of Skellam distribution to improve the fit on draw probabilities, namely 
\begin{equation}
\left\{
\begin{aligned}
\tilde{P}(N(1) = 0) &= p + (1-p) P(N(1) = 0)\\
\tilde{P}(N(1) = x) &= (1-p) P(N(1) = x) \qquad \text{if }x\neq 0.
\end{aligned}
\right.
\end{equation}

Here $0<p<1$ is an inflation factor and $\tilde{P}$ denotes the inflated probabilities. We also consider another type of inflation here
\begin{equation}
\left\{
\begin{aligned}
\tilde{P}(N(1) = 0) &= (1+\theta) P(N(1) = 0)\\
\tilde{P}(N(1) = x) &= (1-\gamma) P(N(1) = x) \qquad \text{if }x\neq 0
\end{aligned}
\right.
\end{equation}
where $\theta$ is the inflation factor and $P(N(1) = 0) = \gamma/(\gamma+\theta)$.

Both types of inflation factors have the corresponding interpretation regarding the bookmakers' way of setting odds. With the first type of factor, the bookmakers generate two different set of probabilities, one specifically for the draw probability (namely the inflation factor $p$) and the other for all the outcomes using the Skellam model. The ``market vig" for all the outcomes is a constant. With the second type, the bookmakers use the Skellam model to generate the probabilities for all the outcomes. Then they apply a larger ``market vig" for draws than others. \cite{yates1982} also point out the ``collapsing" tendency in forecasting behavior, whereby the bookmakers are inclined to report forecasts of 50\% when they feel they know little about the event. In Figure \ref{18games} right panel, we see that the Skellam implied $\log(odds)$ has a heavier right tail than the market implied $\log(odds)$. This effect results from the overestimation of extreme outcomes, which in turn is due to market microstructure effect due to the market ``vig". 

\begin{figure}[ht!]
\centering 
\includegraphics[width=7in, height=3.5in]{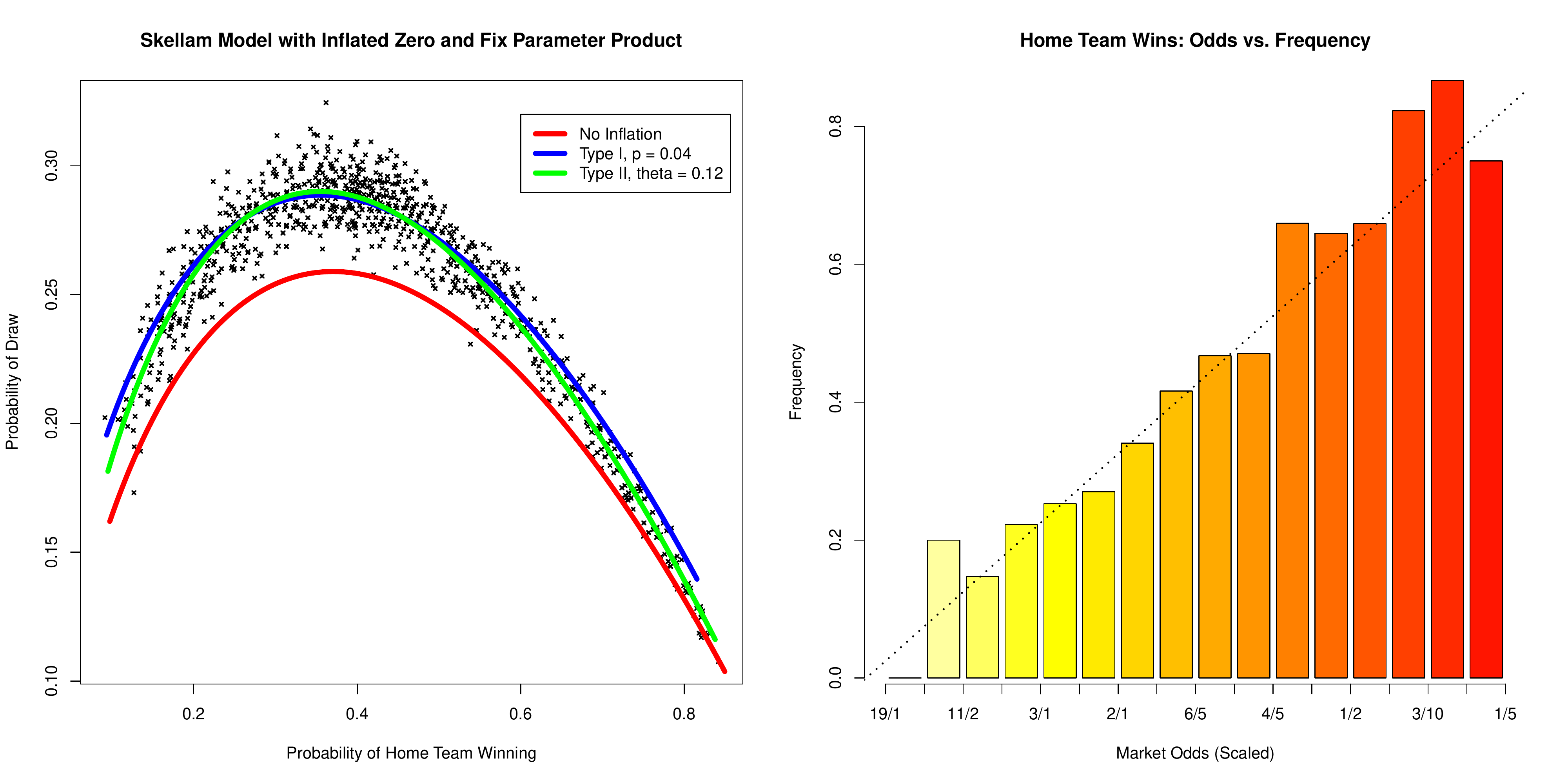} 
\caption{Left: Market implied probabilities of win and draw. The fitted curves are Skellam implied probabilities with fixed $\lambda^A\lambda^B = 1.8$. Right: Market odds and result frequency of home team winning. 1520 EPL games from 2012 to 2016 are used. The dashed line represents: Frequency = Market Implied Probability}
\label{inflation}
\end{figure}
To assess the out-of-sample predictive ability of the Skellam model, we analyze the market (win, lose, draw) odds for 1520 EPL games (from 2012 to 2016, 380 games per season). However, the sample covariance of the end of game scores,$N_A(1)$ and $N_B(1)$, is close to 0. If we assume parameters stay the same, then the estimates are $\hat\lambda^A=1.5$ and $\hat\lambda^B=1.2$. Since the probabilities of win, lose and draw sum to 1, we only plot the market implied probabilities of win and draw.  In Figure \ref{inflation} left panel, the draw probability is nearly a non-linear function of the win probability. To illustrate our model, we set the value of $\lambda^A\lambda^B = 1.5 \times 1.2 = 1.8$ and plot the curve of Skellam implied probabilities (red line). We further provide the inflated Skellam probabilities (blue line for the first type and green line for the second type). As expected, the non-inflated Skellam model (red line) underestimates the draw probabilities while the second type inflated Skellam model (green line) produces the better fit. We also group games by the market implied winning probability of home teams $P(N(1)>0)$: (0.05,0.1], (0.1,0.15], $\cdots$, (0.8,0.85]. We calculate the frequency of home team winning for each group. In Figure \ref{inflation} right panel, the barplot of frequencies (x-axis is regarding scaled odds) shows that the market is efficient, i.e., the frequency is close to the corresponding market implied probability and our Skellam model is calibrated to the market outcome for this dataset.

\subsection{Time-Varying Extension}
One extension that is clearly warranted is allowing for time-varying $\{\lambda^A_t, \lambda^B_t\}$ where the Skellam model is re-calibrated dynamically through updated market odds during the game. We use the current $\{\lambda^A_t, \lambda^B_t\}$ to project possible results of the match in our Skellam model. Here $\{\lambda^A_t, \lambda^B_t\}$ reveal the market expectation of scoring difference for both teams from time $t$ to the end of the game as the game progresses. Similar to the martingale approach of \cite{Polson:2015ira}, $\{\lambda^A_t, \lambda^B_t\}$ reveal the best prediction of the game result. From another point of view, this approach is the same as assuming homogeneous rates for the rest of the game.

An alternative approach to time-varying $\{\lambda^A_t, \lambda^B_t\}$ is to use a Skellam regression with conditioning information such as possession percentages, shots (on goal), corner kicks, yellow cards, red cards, etc. We would expect jumps in the $\{\lambda^A_t, \lambda^B_t\}$ during the game when some important events happen. A typical  structure takes the form
\begin{equation}
\left\{
\begin{aligned}
\log(\lambda^A_t) &=& \alpha_A + \beta_A X_{A,t-1} \\
\log(\lambda^B_t) &=& \alpha_B + \beta_B X_{B,t-1},
\end{aligned}
\right.
\end{equation}
estimated using standard log-linear regression. 

Our approach relies on the betting market being efficient so that the updating odds should contain all information of game statistics. Using log differences as the dependent variable is another alternative with a state space evolution. \cite{Koopman2014} adopt stochastically time-varying densities in modeling the Skellam process. \cite{Barndorff-Nielsen2012a} is another example of the Skellam process with different integer valued extensions in the context of high-frequency financial data. Further analysis is required, and this produces a promising area for future research.

\section{Example: Everton vs West Ham (3/5/2016) }

We collect the real-time online betting odds data from {\tt ladbrokes.com} for an EPL game between Everton and West Ham on March 5th, 2016. By collecting real-time online betting data for every 10-minute interval, we can show the evolution of betting market prediction on the final result. We do not account for the overtime for both 1st half and 2nd half of the match and focus on a 90-minute game. 

\subsection{Implied Skellam Probabilities}
\begin{table}[ht!]
\centering
\begin{tabular}{@{}ccccccc@{}}
\toprule
Everton \textbackslash West Ham & 0 & 1 & 2 & 3 & 4 & 5 \\ \midrule
0 & 11/1 & 12/1 & 28/1 & 66/1 & 200/1 & 450/1 \\
1 & 13/2 & 6/1 & 14/1 & 40/1 & 100/1 & 350/1 \\
2 & 7/1 & 7/1 & 14/1 & 40/1 & 125/1 & 225/1 \\
3 & 11/1 & 11/1 & 20/1 & 50/1 & 125/1 & 275/1 \\
4 & 22/1 & 22/1 & 40/1 & 100/1 & 250/1 & 500/1 \\
5 & 50/0 & 50/1 & 90/1 & 150/1 & 400/1 &  \\ 
6 & 100/1 & 100/1 & 200/1 & 250/1 & &  \\ 
7 & 250/1 & 275/1 & 375/1 & & &  \\ 
8 & 325/1 & 475/1 & & & &  \\ \bottomrule
\end{tabular}
\caption{Original odds data from Ladbrokes before the game started\label{Table1}}
\end{table}
Table \ref{Table1} shows the raw data of odds right the game. We need to transform odds data into probabilities. For example, for the outcome 0-0, 11/1 is equivalent to a probability of 1/12. Then we can calculate the marginal probability of every score difference from -4 to 5. We neglect those extreme scores with small probabilities and rescale the sum of event probabilities to one.

\begin{figure}[htb!]
\centering 
\includegraphics[scale=0.6]{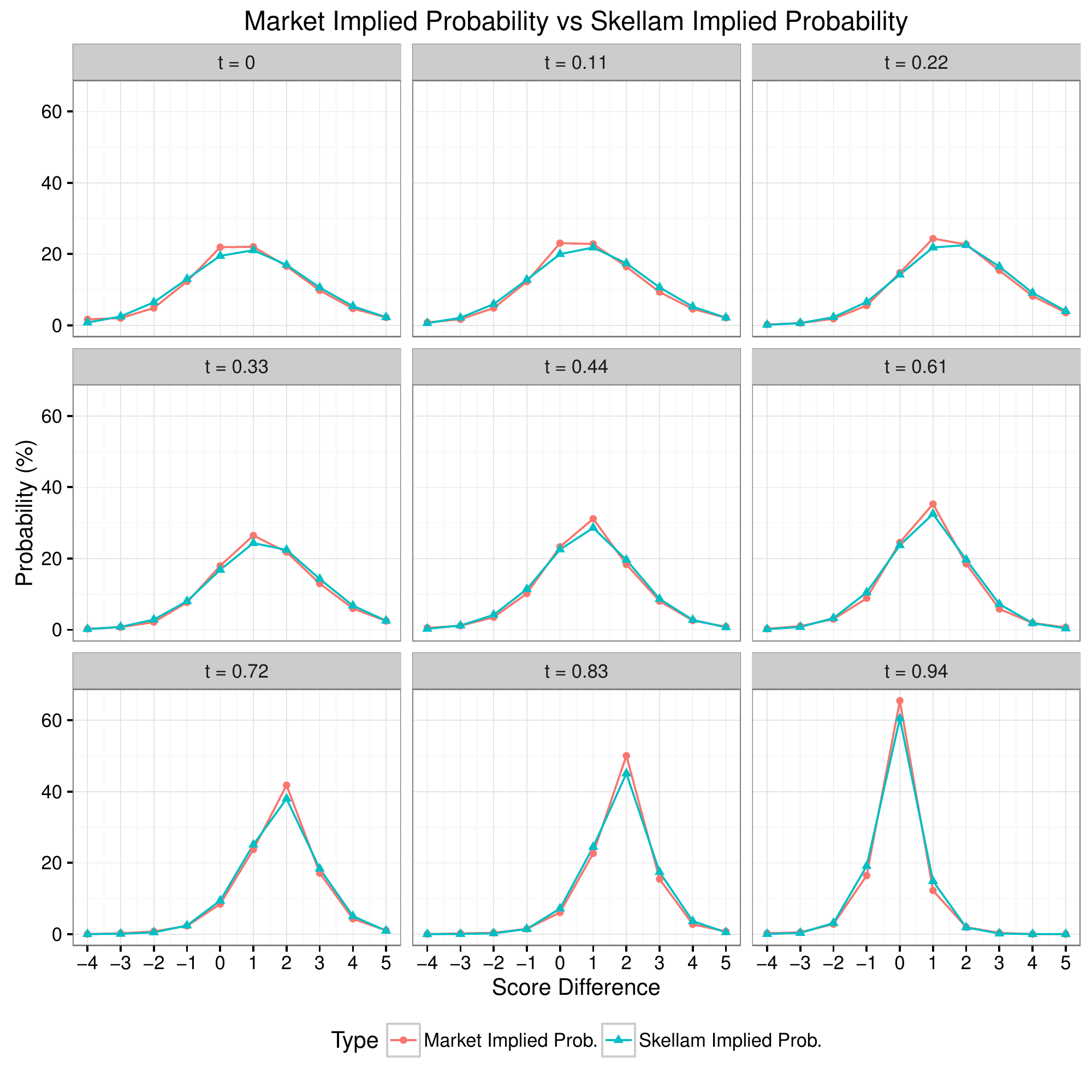} 
\caption{Market implied probabilities versus the probabilities estimated by the model at different time points, using the parameters given in Table \ref{lambda} \label{comparison}.}
\end{figure}

In Figure \ref{comparison}, the probabilities estimated by the model are compared with the market implied probabilities. As we see, during the course of the game, the Skellam assumption suffices to approximate market expectation of score difference distribution. This set of plots is evidence of goodness-of-fit the Skellam model.

\begin{table}[ht!]
\centering
\begin{tabular}{c c c c c c c c c c c}
\toprule
Score difference&-4&-3&-2&-1&0&1&2&3&4&5\\
\midrule
Market Prob. (\%)& 1.70 & 2.03 & 4.88 &12.33& 21.93 &22.06 &16.58  &9.82  &4.72  &2.23\\
Skellam Prob.(\%)& 0.78 & 2.50 & 6.47 & 13.02 & 19.50 & 21.08 & 16.96 & 10.61  & 5.37 & 2.27\\
\bottomrule
  \end{tabular}
  \caption{Market implied probabilities for the score differences versus Skellam implied probabilities at different time points. The estimated parameters $\hat\lambda^A=2.33$, $\hat\lambda^B=1.44.$\label{Table2}}
\end{table}

Table \ref{Table2} shows the model implied probability for the outcome of score differences before the game, compared with the market implied probability. As we see, the Skellam model appears to have longer tails.  Different from independent Poisson modeling in \cite{Dixon:1997jc}, our model is more flexible with the correlation between two teams. However, the trade-off of flexibility is that we only know the probability of score difference instead of the exact scores.

\begin{figure}[ht]
\centering 
\includegraphics[scale=0.45]{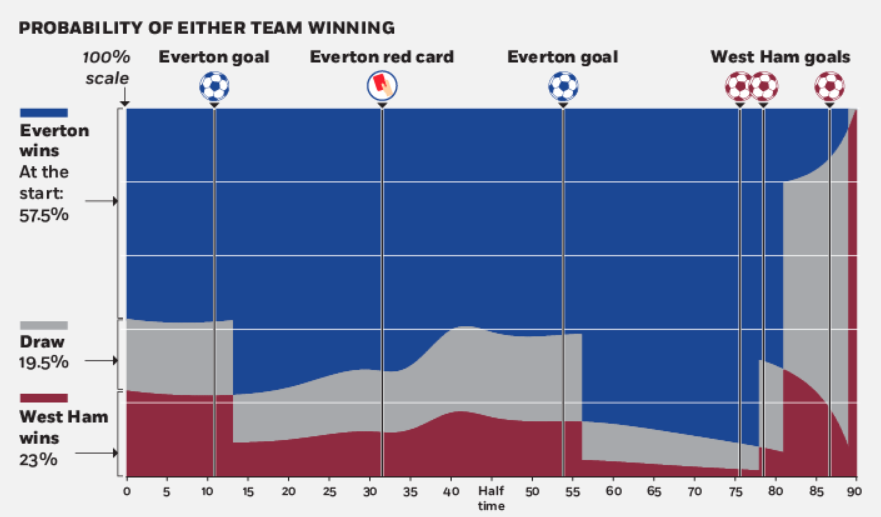} 
\caption{The betting market data for Everton and West Ham is from {\tt ladbrokes.com}. Market implied probabilities (expressed as percentages) for three different results (Everton wins, West Ham wins and draw) are marked by three distinct colors, which vary dynamically as the game proceeds. The solid black line shows the evolution of the implied volatility (defined in Section \ref{IV}). The dashed line shows significant events in the game, such as goals and red cards. Five goals in this game are 13' Everton, 56' Everton, 78' West Ham, 81' West Ham and 90' West Ham.\label{Figure2}}
\end{figure}

Finally, we can plot these probability paths in Figure \ref{Figure2} to examine the behavior of the two teams and represent the market predictions on the final result. Notably, we see the probability change of win/draw/loss for important events during the game: goals scoring and a red card penalty. In such a dramatic game, the winning probability of Everton gets raised to 90\% before the first goal of West Ham in 78th minutes. The first two goals scored by West Ham in the space of 3 minutes completely reverses the probability of winning. The probability of draw gets raised to 90\% until we see the last-gasp goal of West Ham that decides the game.

\subsection{How the Market Forecast Adapts} \label{IV}

A natural question arises to how does the market odds (win, lose, draw and actual score) adjust as the game evolves. This is similar to option pricing where Black-Scholes model uses its implied volatility to show how market participants' beliefs change. Our Skellam model mimics its way and shows how the market forecast adapts to changing situations during the game. See \cite{Merton:1976ge} for references of jump models.

Our work builds on \cite{Polson:2015ira} who define the implied volatility of a NFL game. For an EPL game, we simply define the implied volatility as $\sigma_{IV,t} = \sqrt{\lambda^A_t + \lambda^B_t}$. As the market provides real-time information about $\lambda^A_t$ and $\lambda^B_t$, we can dynamically estimate $\sigma_{IV,t}$ as the game proceeds. Any goal scored is a discrete Poisson shock to the expected score difference (Skellam process) between the teams, and our odds implied volatility measure will be updated.

Figure \ref{Figure2} plots the path of implied volatility throughout the course of the game. Instead of a downward sloping line, we see changes in the implied volatility as critical moments occur in the game. The implied volatility path provides a visualization of the conditional variation of the market prediction for the score difference. For example, when Everton lost a player by a red card penalty at 34th minute, our estimates $\hat\lambda^A_t$ and $\hat\lambda^B_t$ change accordingly. There is a jump in implied volatility and our model captures the market expectation adjustment about the game prediction. The change in $\hat\lambda_A$ and $\hat\lambda_B$ are consistent with the findings of \cite{Vecer2009} where the scoring intensity of the penalized team drops while the scoring intensity of the opposing team increases. When a goal is scored in the 13th minute, we see the increase of $\hat\lambda^B_t$ and the market expects that the underdog team is pressing to come back into the game, an effect that has been well-documented in the literature. Another important effect that we observe at the end of the game is that as goals are scored (in the 78th and 81st minutes), the markets expectation is that the implied volatility increases again as one might expect. 

\begin{figure}[ht!]
\centering 
\includegraphics[scale=0.5]{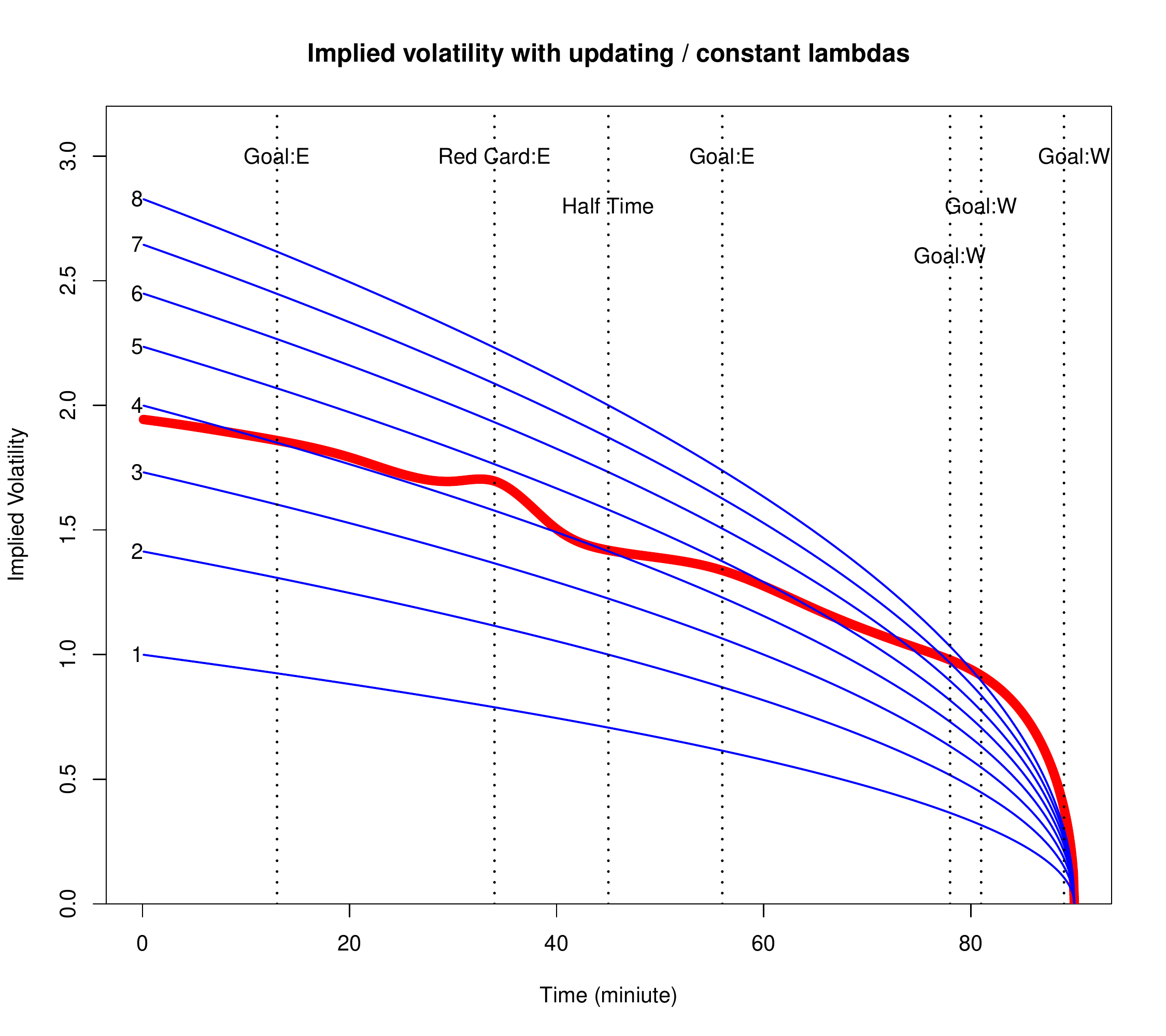} 
\caption{Red line: the path of implied volatility throughout the game, i.e., $\sigma_{t}^{red} = \sqrt{\hat\lambda^A_t+\hat\lambda^B_t}$. Blue lines: the path of implied volatility with constant $\lambda^A+\lambda^B$, i.e., $\sigma_{t}^{blue} = \sqrt{(\lambda^A+\lambda^B)*(1-t)}$. Here $(\lambda^A+\lambda^B) = 1, 2, ..., 8$. \label{ivcompare}}
\end{figure}

\begin{table}[ht!]
\centering
\begin{tabular}{c c c c c c c c c c c c}
\toprule
t & 0 & 0.11 & 0.22 & 0.33 & 0.44 & 0.50 & 0.61 & 0.72 & 0.83 & 0.94 & 1\\
\midrule
$\hat\lambda^A_t/(1-t)$ & 2.33 & 2.51 & 2.53 & 2.46 & 1.89 & 1.85 & 2.12 & 2.12 & 2.61 &  4.61 &  0\\

$\hat\lambda^B_t/(1-t)$ & 1.44 & 1.47 & 1.59 & 1.85 & 2.17 & 2.17 & 2.56 & 2.90 & 3.67 & 5.92 & 0\\

\midrule

$(\hat\lambda^A_t+\hat\lambda^B_t)/(1-t)$ & 3.78  & 3.98  & 4.12  & 4.31  & 4.06  & 4.02  & 4.68  & 5.03  & 6.28 & 10.52  &0\\
\midrule
$\sigma_{IV,t}$ & 1.94 & 1.88 & 1.79 & 1.70 & 1.50 & 1.42 & 1.35 & 1.18 & 1.02 & 0.76 & 0\\
\bottomrule
\end{tabular}
\caption{The calibrated $\{\hat\lambda^A_t, \hat\lambda^B_t\}$ divided by $(1-t)$ and the implied volatility during the game. $\{\lambda^A_t, \lambda^B_t\}$  are expected goals scored for rest of the game. The less the remaining time, the less likely to score goals. Thus $\{\hat\lambda^A_t, \hat\lambda^B_t\}$ decrease as $t$ increases to 1. Diving them by $(1-t)$ produces an updated version of $\hat\lambda_{0}$'s for the whole game, which are in general time-varying (but not decreasing necessarily).\label{lambda}}
\end{table}

Figure \ref{ivcompare} compares the updating implied volatility of the game with implied volatilities of fixed $(\lambda^A+\lambda^B)$. At the beginning of the game, the red line (updating implied volatility) is under the "($\lambda^A+\lambda^B=4)$"-blue line; while at the end of the game, it's above the "($\lambda^A+\lambda^B=8)$"-blue line. As we expect, the value of $(\hat\lambda^A_t + \hat\lambda^B_t)/(1-t)$ in Table \ref{lambda} increases throughout the game, implying that the game became more and more intense and the market continuously updates its belief in the odds.


\section{Discussion}
The goal of our analysis is to provide a probabilistic methodology for calibrating real-time market odds for the evolution of the score difference for a soccer game.Rather than directly using game information, we use the current odds market to calibrate a Skellam model to provide a forecast of the final result. To our knowledge, our study is the first to offer an interpretation of the betting market and to show how it reveals the market expectation of the game result through an implied volatility.  One area of future research is studying the index betting. For example, a soccer game includes total goals scored in match and margin of superiority (see \cite{Jackson:1994gj}). The latter is the score difference in our model, and so the Skellam process directly applies.

Our Skellam model is also valid for low-scoring sports such as baseball, hockey or American football with a discrete series of scoring events. For NFL score prediction, \cite{baker2013} propose a point process model that performs as well as the betting market. On the one hand, our model has the advantage of implicitly considering the correlation between goals scored by both teams but on the other hand, ignores the sum of goals scored.  For high-scoring sports, such as basketball, the Brownian motion adopted by \cite{Stern:1994hj} is more applicable. \cite{Rosenfeld:1000a} provides an extension of the model that addresses concerns of non-normality and uses a logistic distribution to estimate the relative contribution of the lead and the remaining advantage. Another avenue for future research, is to extend the Skellam model to allow for the dependent jumpiness of scores which is somewhere in between these two extremes (see \cite{Glickman:2012dt}, \cite{Polson:2015ira} and \cite{Rosenfeld:1000a} for further examples.)

Our model allows the researcher to test the inefficiency of EPL sports betting from a statistical arbitrage viewpoint. More importantly, we provide a probabilistic approach for calibrating dynamic market-based information. \cite{Camerer:1989dc} shows that the market odds are not well-calibrated and that an ultimate underdog during a long losing streak is underpriced on the market. \cite{Golec:1991cd} test the NFL and college betting markets and find bets on underdogs or home teams win more often than bets on favorites or visiting teams. \cite{Gray:1997gz} examine the in-sample and out-of-sample performance of different NFL betting strategies by the probit model. They find the strategy of betting on home team underdogs averages returns of over 4 percent, over commissions. In summary, a Skellam process appears to fit the dynamics of EPL soccer betting very well and produces a natural lens to view these market efficiency questions.

\newpage
\bibliography{EPL_0918.bib}
\end{document}